\documentclass[aps,twocolumn,nofootinbib,showpacs, prd]{revtex4-2}
\usepackage{graphicx}
\usepackage{amsfonts}
\usepackage{amssymb}
\usepackage{amsbsy}
\usepackage{amsmath}
\usepackage{latexsym}
\usepackage{natbib}
\usepackage{bm}
\usepackage{color}
\usepackage{float}

\begin{document}
\title{Curvature-driven Acceleration and a notion of Vacuum in Nash Theory}
\author{Soumya Chakrabarti\footnote{soumya.chakrabarti@vit.ac.in}}
\affiliation{School of Advanced Sciences \\
Vellore Institute of Technology \\ 
Tiruvalam Rd, Katpadi, Vellore, Tamil Nadu 632014 \\
India}
\author{Soumya Bhattacharya\footnote{soumya557@bose.res.in}, Rabin Banerjee\footnote{rabin@bose.res.in}, Amitabha Lahiri \footnote{amitabha@bose.res.in}}
\affiliation{Department of Theoretical Sciences \\
S. N. Bose National Centre for Basic Sciences \\
JD Block, Sector-III, Salt Lake City, Kolkata - 700 106 \\
India}


\date{\today}

\begin{abstract}
We discuss the cosmological implications in the \textit{Nash theory} of gravity, where the field equations are derived from a lagrangian quadratic in Ricci invariants. A spatially homogeneous exact solution is found which can serve as a cosmological toy model. The implications of this model (departures from a standard $\Lambda$CDM cosmology) are compared with standard observational expectations, using a Markov chain Monte Carlo simulation of $JLA + OHD + BAO$ data sets. We propose that the Nash vacuum dynamics can be imagined as the effective dynamics of a growing vacuum in General Relativity, written as a time-evolving scalar field. The resulting mild evolution of the masses of fundamental particles is found to be within the observed scale of variation measured through the molecular absorption spectra of a series of Quasars.  
\end{abstract}

\maketitle

\section{Introduction}
Einstein's General Theory of Relativity (GR) allows us to put space-time, matter and gravity within one bracket; a single mathematical framework on the macroscopic level. Not only the theory stands out for its elegance, it keeps on passing experimental tests over the years; the most recent on the list being the detection of gravitational waves \cite{Ligo} and the image of black-hole shadows captured by Event Horizon Telescope \cite{EHT}. There are albeit a few remaining riddles within the theory related to patches of observational inconsistencies \cite{riess, paddy, maor, upadhye, zlatev, adel} and also, some limitations in probing the theory in higher curvature regime. The latter problem is rooted in the incompatibility of GR and quantum field theory, however, higher curvature corrections are often considered as a possible solution. Both of these issues leave open the scope of working out viable modified theories of gravity. The particular theory of gravity proposed by John. Nash Jr. \cite{Nash_lec} (Nash Theory from here onwards) is one of a kind in this class. \\

The signature of Nash theory is a specific choice of gravitational action, quadratic in the Ricci scalar and Ricci tensor. The action has some resemblance with string inspired theories of gravity and also a few characteristic departures \cite{string}. For instance, it has no terms linear in Ricci scalar in the action and the comparative weighting of the two quadratic terms $R^{\alpha\beta}R_{\alpha\beta}$ and $R^{2}$ are different from a Lovelock invariant \cite{lovelock}. This indicates higher order field equations. This is probably one of the reasons for which the theory is relatively less explored (apart from a few attempts \cite{Gron, Thai1, Thai2}), despite accommodating some remarkable features. For instance, the theory is a special case of the renormalizable Stelle Gravity \cite{Stelle}, and therefore a possible candidate for quantum gravity theories. Moreover, the scalar equation of Nash Theory in a four dimensional vacuum is a wave equation and it can enable a wider variety of gravitational waves compared to GR. After all, it is not wrong to expect new solutions governing the nature of space-time geometry from a set of higher order field equations. On this note, we focus on vacuum cosmological solutions of the Nash field equations. The manner in which an energy-momentum distribution can enter the Nash theory is a different question altogether and we avoid that question in this work. We find an exact solution of the spatially flat cosmological equations in vacuum. We also show that a Nash vacuum may be rendered as a combination of two parts : vacuum Einstein field equations plus an exotic self-interacting scalar field. This \textit{vacuum} field has a symmetry breaking self-interaction (like Higgs) and allows a mildly varying vacuum expectation value (\textit{vev}). This leads to variations of quark masses and written in a quantified manner through a variation of proton-to-electron mass ratio $\mu$. We show that these variations are on a desired scale as predicted by the molecular absorption spectra of a series of Quasars. \\

In section \ref{Nash_basics} we briefly review the action and the field equations of Nash theory. We introduce the spatially flat FRW equations in the same section and solve them in section \ref{Nash_cosmology}. Evolution of the Hubble parameter and kinematic quantities governing the late-time acceleration of the universe are discussed in the same section. In section \ref{vacuum_dynamics} we discuss the structure of an envisaged vacuum scalar field and the scale of variation of different fundamental couplings. We conclude the manuscript in section \ref{conclusions}. 

\section{Nash equations for Vacuum} \label{Nash_basics}
The Nash action is written as
\begin{equation}\nonumber\nonumber
S = \int d^4x \sqrt{-g} \left[2R^{\alpha\beta}R_{\alpha\beta} - R^{2}\right].
\end{equation}
We can see straightaway that there is no term linear in Ricci scalar in the action, i.e., the standard Einstein-Hilbert part is switched off. The field equations are derived by taking a metric variation and written as
\begin{equation}
N^{\mu\nu} \equiv \Box G^{\mu\nu} + G^{\alpha\beta}\left(2R_{\alpha~~\beta}^{~\mu~~~\nu} - \frac{1}{2}g^{\mu\nu}R_{\alpha\beta}\right) = 0,
\label{Nasheq}
\end{equation}
which will henceforth be called the Nash equations. The $\Box$ is the d'Alembertian and $G^{\mu \nu} = R^{\mu \nu} - \frac{1}{2} g^{\mu \nu} R$ is the standard Einstein tensor. Clearly, any solution of the vacuum Einstein equations is also a solution of these~\cite{Nash_lec}. It is also easy to see that any space of arbitrary constant curvature, i.e. satisfying $R_{\mu\nu} = \Lambda g_{\mu\nu}\,,$ is also a solution of Eq.~(\ref{Nasheq}).
However, these equations can provide for a wider variety of solutions due to their nature, involving higher curvature terms in a very specific way. One peculiarity of gravitational theories with higher curvature terms is that it yields equations of motion of higher-order in derivatives of the metric. Due to Ostrogradsky's theorem  it is a well-known fact that non-degenerate Lagrangians containing second (or higher) derivatives of a field give rise to some classical instabilities. A further analysis in \cite{Woodard} have shown specifically that the presence of the term $R_{\mu\nu} R^{\mu \nu}$ gives rise to two unconstrained instabilities per space point. But then again one can argue that Ostrogradsky instability is only a classical instability and it does not necessarily mean a quantum instability as argued in \cite{Donoghue}. Moreover, it has been suggested by these authors that these type of instabilities can be removed by quantum physics. So keeping this in mind we can venture further and see what the Nash theory predicts. We take a spatially flat FRW spacetime as
\begin{equation}
ds^{2} = -dt^{2} + a(t)^{2}\left(dr^{2} + r^{2}d\Omega^{2}\right).
\end{equation}
The components of the Nash tensor take the form
\begin{equation}\label{nash1}
N^{tt} = -3\left(2\ddot{H}H + 6\dot{H}H^{2} - \dot{H}^{2}\right),
\end{equation}
and
\begin{equation}\label{nash2}
N^{rr} = 2\overset{\ldots}{H} + 12\ddot{H}H + 9\dot{H}^{2} + 18\dot{H}H^{2}.
\end{equation}

All the components are written as a function of $H = \frac{\dot{a}}{a}$, the Hubble function. We can then write the trace of the Nash tensor ($N = g_{\mu \nu} N^{\mu \nu}$) as
\begin{equation}\label{trace}
{\textit N} = tr(N) = 6\left(\overset{\ldots}{H} + 7\ddot{H}H + 4\dot{H}^{2} + 12\dot{H}H^{2}\right).
\end{equation}

We can further simplify the equations by introducing the variable \cite{Gron}
\begin{equation}
y = H^{1/2},
\end{equation}
which leads to the equations
\begin{eqnarray}
&& N^{tt} = -12y^{3}\left(\ddot{y} + 3y^{2}\dot{y}\right) = 0, \\&&
N^{rr} = 4\left(\overset{\ldots}{y}y + 6\ddot{y}y^{3} + 3\dot{y}\ddot{y} + 9y^{5}\dot{y} + 15\dot{y}^{2}y^{2}\right) = 0.
\end{eqnarray}

With a simple manipulation we rewrite these as
\begin{eqnarray} \label{nashmanip1}
&& \frac{d}{dt}\left(\dot{y} + y^{3}\right) = 0, \\&&\label{nashmanip2}
y\frac{d^{2}}{dt^{2}}\left(\dot{y} + y^{3}\right) + \frac{3}{2}\frac{d}{dt}\left\lbrace\left(\dot{y} + y^{3}\right)^{2}\right\rbrace = 0.
\end{eqnarray}
It is interesting to note that only one independent component $\frac{d}{dt}\left(\dot{y} + y^{3}\right)$ dictates the structure of the field equations for a Nash vacuum. From Eq. (\ref{nash1}), this component is identified as
\begin{equation}\label{vaccnash1}
-3\left(2\ddot{H}H + 6\dot{H}H^{2} - \dot{H}^{2}\right) = 0.
\end{equation}
This feature of one independent component Eq. (\ref{vaccnash1}) dictating dynamics in gravitational field is a unique feature of Nash gravity. Such an equation can not be found in standard GR. It is already known that Nash equations accommodate two solutions of standard GR, the Milne and the de-Sitter metric \cite{Gron}, however, these do not support a late-time acceleration of the universe. In the next section we focus on solving the spatially flat Nash vacuum equations for a solution that can describe the present acceleration of the universe, preceded by a deceleration.

\section{Exact Solution and Cosmological Analysis} \label{Nash_cosmology}
We find that the independent Nash equation can be written in terms of purely kinematic parameters involved in cosmology, in particular, the deceleration ($q$) and the jerk parameter ($j$). The standard parameters, Hubble, deceleration, jerk and statefinder are defined respectively as
\begin{eqnarray}
H & = & \frac{\dot{a}}{a} \label{eq:1.1} \\
q & = & -\frac{\ddot{a}a}{\dot{a}^2} = -\frac{\dot H}{H^2}-1, \\ \label{eq:eq1.2}
j & = & \frac{\ddot H}{H^3}+3\frac{\dot H}{H^2}+1\label{eq:eq1.3} \\
s & = & \frac{j-1}{3\left(q-\frac{1}{2}\right)}. \label{eq:eq1.4}
\end{eqnarray}

We write Eq. (\ref{vaccnash1}) as a linear combination of $j$ and $q$ as follows
\begin{equation}\label{jqrelation}
-2(j-1) + (q+1)^{2} = 0.
\end{equation}
\begin{figure}[t!]
\begin{center}
\includegraphics[angle=0, width=0.40\textwidth]{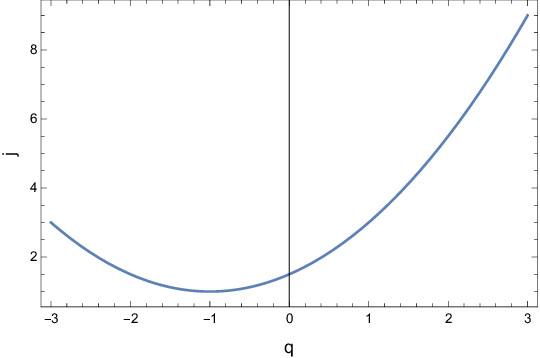}
\includegraphics[angle=0, width=0.40\textwidth]{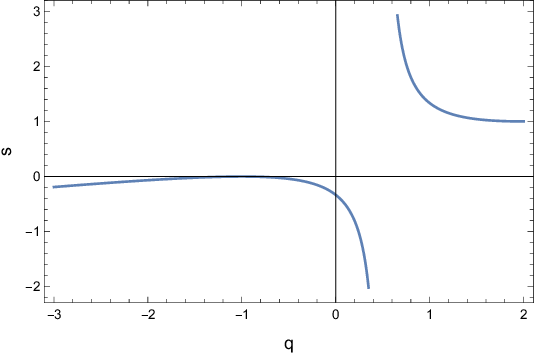}
\caption{Evolution of jerk ($j$) and statefinder ($s$) as a function of the deceleration $q$.}
\label{jqplots}
\end{center}
\end{figure}
This is already a new result that puts significant constraints on the allowed evolution of the universe. We note below a few non-trivial features and resulting speculations from Eq.~(\ref{jqrelation}), even before solving for $H$.
\begin{enumerate}
\item {It is not possible for the jerk parameter to have a negative value, anywhere during the expansion of the Universe described by a Nash vacuum. This restriction is not found in standard GR.}
\item {The statefinder parameter becomes
\begin{equation}\label{special}
s = \frac{(q+1)^{2}}{6\left(q-\frac{1}{2}\right)}.
\end{equation}
For all $q < 0$, the statefinder is negative. This indicates that a Nash vacuum allows the statefinder parameter to attain only negative values during an epoch of acceleration. Only for $q > \frac{1}{2}$, the statefinder can be positive. This restriction is not found in standard GR.}
\item {At any {\textit `critical point'} where the universe goes through a transition from deceleration into acceleration or vice versa, $q = 0$, which indicates $j = \frac{3}{2}$ and $s = -\frac{1}{3}$. Therefore, the point of transition(s) is fixed on the parameter space at the outset.}
\item {By definition, at $q = \frac{1}{2}$, there is a divergence of the statefinder parameter (See Eq. (\ref{eq:eq1.4})). Therefore, the deceleration can take a value either in the $q < \frac{1}{2}$ domain or in the $q > \frac{1}{2}$ domain. The former one seems more plausible, since the deceleration is supposed to have a smooth transition from positive into negative domain and vice versa during the evolution of the universe, without any discontinuity.}
\end{enumerate}

Altogether, a Nash vacuum solution seems to provide a constrained cosmological dynamics, with specific allowed values of kinematic parameters in different phases of the universe. However, this is a new spatially homogeneous exact solution and can serve as a simple cosmological toy model. Most of the available alternative models of late-time cosmology closely follow and reiterate the dynamics of standard $\Lambda$CDM cosmology. However, it is an interesting case when a simple modified theory of gravity with a specific restriction on the kinematic parameters as in Eq. (\ref{jqrelation}) can produce an accelerating solution even without any matter. First, we write Eq. (\ref{vaccnash1}) as an equation whose arguments are redshift rather than cosmic time
\begin{equation}
\frac{H''}{H} + \frac{1}{2}\frac{H'^{2}}{H^{2}} - \frac{2H'}{Hx} = 0,
\end{equation}
where a primes indicates differentiation with respect to $x=(1+z)$.

\begin{figure}[t!]
\begin{center}
\includegraphics[angle=0, width=0.52\textwidth]{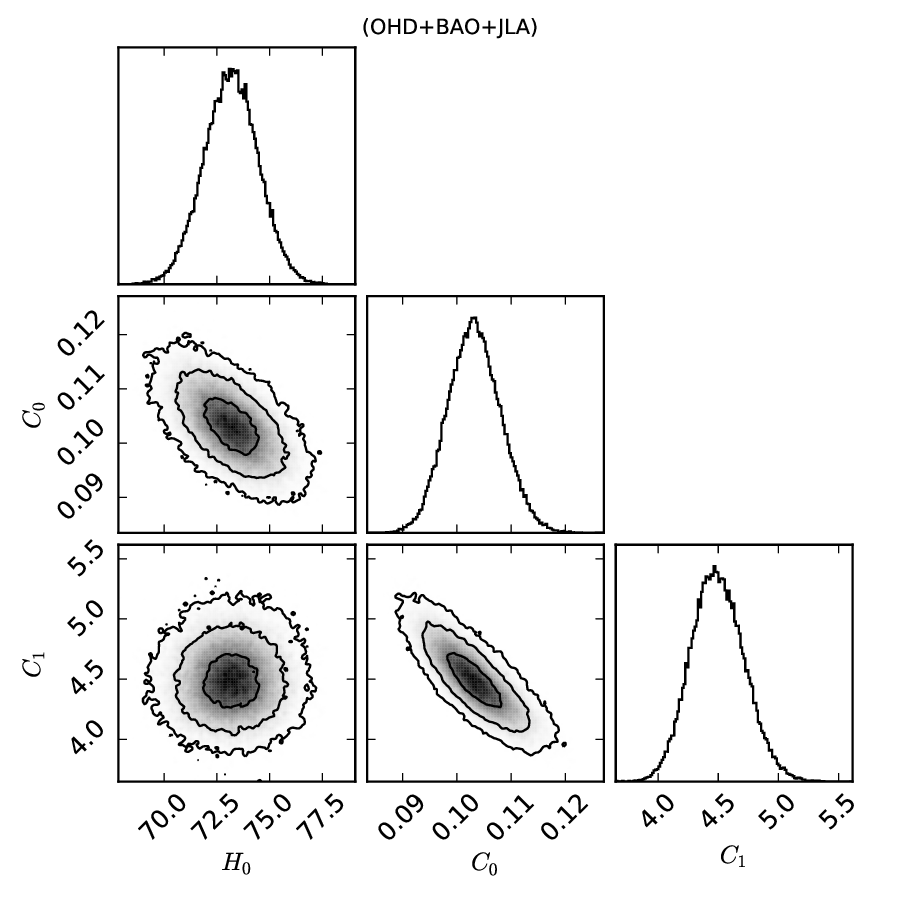}
\caption{Parameter Space Confidence Contours showing estimation of the uncertainty, the best fit and the likelihood analysis of parameters (combination of data from OHD+JLA+BAO).}
\label{Modelcontour}
\end{center}
\end{figure}

\begin{table*}[]
\caption{{\small Best Fit Parameter values of three parameters : (i) Dimensionless Hubble $h_{0}$, (ii) $C_{0}$ and (iii) $C_{1}$.}}\label{resulttable}
\begin{center}
{\centering
\begin{tabular}{l  c  c  c  c  c  c  c}
\hline 
 & \multicolumn{1}{c}{$H_0$} & \multicolumn{1}{c}{$C_{0}$} & \multicolumn{1}{c}{$C_{1}$} \\
\hline \\ 
$OHD+JLA+BAO$ 	  & $73.2^{+1.2}_{-1.2}$ &$0.103^{+0.005}_{-0.005}$ & $4.5^{+0.2}_{-0.2}$ & \\ \, \\
\hline
\end{tabular}
}
\end{center}
\end{table*}

\begin{figure}[t!]
\begin{center}
\includegraphics[angle=0, width=0.40\textwidth]{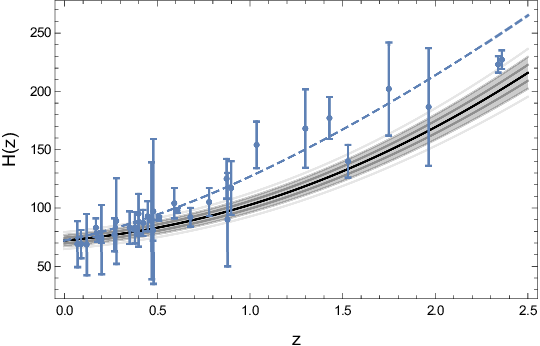}
\caption{Evolution of Hubble Function alongwith observational data points. The thick line is for best fit parameter values and the gray regions are for associated 2$\sigma$ and 3$\sigma$ confidence regions. The dashed curve depicts a corresponding $\Lambda$CDM behavior.}
\label{Hz_data}
\end{center}
\end{figure}

Solving this we write an exact analytical form for the Hubble parameter as
\begin{equation}\label{hubble}
H(z) = H_{0}\left[C_{0}\left\lbrace(1+z)^{3}-1\right\rbrace + C_{0}(1+2C_{1})\right]^{\frac{2}{3}},
\end{equation}
where $H_{0}$ is the value of Hubble parameter at the present time, i.e., at $z = 0$. $C_{0}$ and $C_{1}$ are constants of integration. This is an exact solution and although we do not expect this to match with observational data extremely well, we can indeed establish some basic requirements in view of a consistent late-time cosmology. We need to ensure that we have (i) a consistent present value of the Hubble and deceleration parameter and (ii) evolution curves of the Hubble function and the deceleration fitting in with observations without any discontinuity. An estimation of the model parameters are done to meet these requirements, for which we use the following set of observations
\begin{itemize}
\item {Joint Light Curve Analysis from $SDSS-II$ and $SNLS$ collaborations, resulting in the the Supernova distance modulus data \cite{betoule},}
\item {estimated measurement of Hubble parameter value in the present epoch (OHD) \cite{ohd} and}
\item {the Baryon Acoustic Oscillation (BAO) data from the BOSS collaborations and $6dF$ $Galaxy$ $Survey$ \cite{bao}.}
\end{itemize}
We have used a \textit{Markov Chain Monte Carlo simulation} (MCMC) written in python. The analysis is statistical and the results are shown in the form of confidence contours on the parameter space, as in Fig. \ref{Modelcontour}. The contours point out to the best fit values of three parameters : (i) Dimensionless Hubble $h_{0}$ ($\sim H_{0}/100 km Mpc^{-1} sec^{-1}$), (ii) $C_{0}$ and (iii) $C_{1}$. The best possible values of the three parameters and 1$\sigma$ error estimations are written in Table.~\ref{resulttable}. $H_{0}$, the current value of Hubble parameter is quite consistent with the observations \cite{ohd}. The evolution of $H(z)$ as a function of $z$ is shown in Fig.~\ref{Hz_data}. Observational data points are fitted alongwith the curve which shows sufficient match during the late-times. However, for $z > 1$, the departure from standard $\Lambda$CDM is apparent. \\

\begin{figure}[t!]
\begin{center}
\includegraphics[angle=0, width=0.40\textwidth]{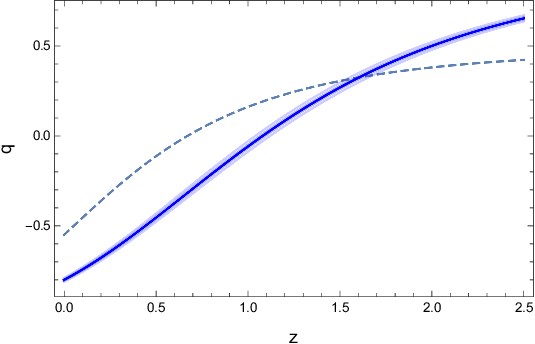}
\includegraphics[angle=0, width=0.40\textwidth]{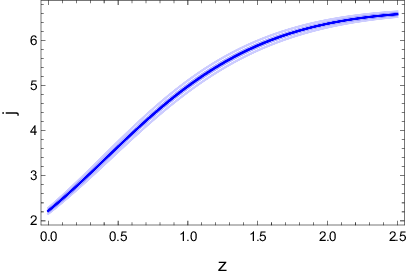}
\caption{Evolution of deceleration and jerk as a function of redshift. The dashed curve depicts a corresponding $\Lambda$CDM behavior of deceleration parameter. Note that the jerk parameter for $\Lambda$CDM is unity.}
\label{qandj_data}
\end{center}
\end{figure}

The curves depicting deceleration and the jerk parameter evolutions are shown in Fig. \ref{qandj_data} and they suggest a departure from standard cosmology. Around $z \sim 0$, i.e., the present time, the deceleration is close to $-0.8$ which indicates a universe expanding at a slightly faster rate than the anticipated acceleration (for standard $\Lambda$CDM cosmology this value is $\sim$ $0.65$). The redshift of transition from deceleration into the acceleration is $z_{t} < 1$ which goes well with observations. For a higher range of redshift, the deceleration becomes positive, indicating a decelerated expansion immediately prior to the present epoch. The jerk parameter is close to $2$ around $z \sim 0$ which is quite different from standard $\Lambda$CDM, for which jerk is equal to $1$. Moreover, the plots in Fig. \ref{jqplots} clearly suggests that a Nash acceleration without matter is quite different from $\Lambda$CDM. Without a matter component, we do not expect a flexibility from this curvature-driven acceleration to accommodate different epochs of matter/radiation domination. It is clear at this point that even without matter the theory is capable of describing a solution. However, special (simple) cases do produce solutions extremely suggestive of an inflation. For instance, Eqs. (\ref{nashmanip1}) and (\ref{nashmanip2}) simultaneously lead to the condition 
\begin{equation}
\dot{y} + y^3 = K,
\end{equation}
where $K$ is a constant. If we take $K=0$ and also the inflationary limit $H \rightarrow \infty$ as $t\rightarrow 0\,,$ we find that this equation gives the solution of the scale factor to be $a(t) \propto t/t_{0}$\,, where $t_{0}$ is the current age of the universe.  This is nothing but a coasting universe solution or Minkowski metric written in an expanding reference frame. Similarly, for $K \neq 0$, a particular solution of the above equation is $a(t) = e^{H(t-t_{0})}$. Separately these solutions can describe inflation, however, the impasse is once again a correct description of matter. The solution we have found is in fact, not found in GR and unique in its kinematic properties.
\begin{figure}[t!]
\begin{center}
\includegraphics[angle=0, width=0.40\textwidth]{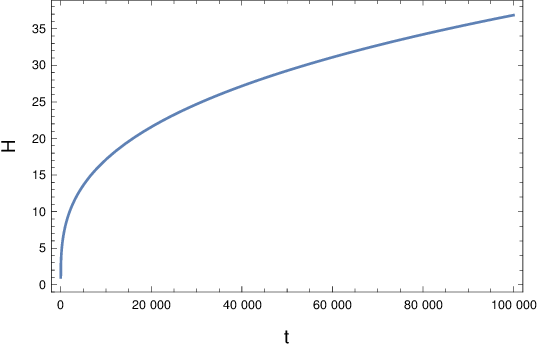}
\caption{Evolution of Hubble as a function of cosmic time.}
\label{H_t}
\end{center}
\end{figure}

\begin{figure}[t!]
\begin{center}
\includegraphics[angle=0, width=0.40\textwidth]{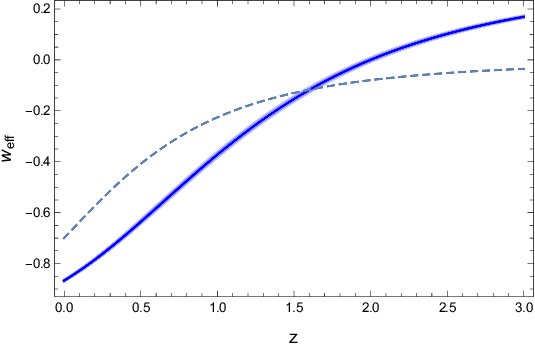}
\caption{Evolution of the effective Equation of State parameter as a function of redshift. The dashed curve depicts a corresponding $\Lambda$CDM behavior.}
\label{weff_data}
\end{center}
\end{figure}

Most of the discussion on cosmic expansion is usually done as a function of look-back time or redshift. It is not easy to articulate the nature of expansion in the remote future, however, one can always solve the Hubble evolution Eq. (\ref{vaccnash1}) as a function of cosmic time. The solution is plotted in Fig. \ref{H_t} and suggests a monotonically increasing nature of Hubble since the inflationary epoch. \\

The effective equation of state $\omega_{eff}$ as shown in Fig. \ref{weff_data}, shows a dark energy dominated acceleration but for $z > 1$ suggests a radiation dominated deceleration. It is important to emphasize here that the effective EOS parameter for a cosmic expansion can be defined as a function of the rate of expansion by first writing
\begin{equation}
w_{eff}=\frac{p_{tot}}{\rho_{tot}},
\end{equation}
where, $\rho_{tot}$ is the total energy density and $p_{tot}$ is the total effective pressure. They are given by the field equations of the theory and are connected to the expansion rate by
\begin{eqnarray}
&&\frac{\rho_{tot}}{\rho_{c0}}=\frac{H^2(z)}{H^2_0},\\&&
\frac{p_{tot}}{\rho_{c0}}=-\frac{H^2(z)}{H^2_0}+\frac{2}{3}\frac{(1+z)H(z)H'(z)}{H^2_0}.
\end{eqnarray}

The present critical density is defined as $\rho_{c0} = 3H^2_0/8\pi G$. In a similar manner, for Nash theory one can define an effective equation of state using the solution of Eq. (\ref{vaccnash1}). In the plots, a dashed curve shows a similarly defined, corresponding $\Lambda$CDM behavior, which confirms the departure of a vacuum Nash cosmology from standard expectations. We believe a proper description of the source of energy-momentum distribution in Nash cosmology can come in as a correction on the level of field equations and provide a lot more observational equivalence. \\

We finish this section with a brief comment on the inclusion of energy-momentum density tensor in Nash equation. Let us consider a simple example of a universe filled with radiation for this sake. Since the trace of the energy-momentum tensor is zero we may proceed by solving the $tr (N) = 0$ equation. From Eq.~(\ref{trace}) this takes the form
\begin{equation}
\left(\overset{\ldots}{H} + 7\ddot{H}H + 4\dot{H}^{2} + 12\dot{H}H^{2}\right) = 0,
\end{equation}
which can also be written as
\begin{equation}
\frac{d^{2}}{dt^{2}}\left\lbrace \dot{H} + 2H^2 \right\rbrace + 3H \frac{d}{dt}\left\lbrace \dot{H} + 2H^2 \right\rbrace = 0.
\end{equation}
We focus on a particular solution obeying
\begin{equation}
\frac{d}{dt}\left\lbrace \dot{H} + 2H^2 \right\rbrace = 0.
\end{equation}
After integration this gives
\begin{equation}
\dot{H} + 2H^2 = 2H_{0}^{2},
\end{equation}
where $H_0$ is a constant. Two exact solutions are possible to consider, for $H_0 = 0$ and $H_0 \neq 0$. For $H_0 = 0$, assuming $a(t)_{t=t_{0}} = 1$, the solution for scale factor can be written as
\begin{equation}
a(t) = \sqrt{1+2(t-t_{0})}.
\end{equation}
This has the same form as the scale factor of a flat radiation dominated FLRW-universe in the Einstein theory. However, calculating the Nash tensor components using this scale factor we find
\begin{equation}
N^{tt} = 0 ~,~ N^{rr} = \frac{30}{a^{12}},
\end{equation}
which can not describe radiation. In other words, $N_{\mu\nu}$ is not always proportional to a $T_{\mu\nu}$ as in GR (for more discussion on this see for instance \cite{Gron}). We can infer this from the non-trivial constraint on deceleration and jerk parameter found in the form of Eq. (\ref{special}). This is an example of acceleration without matter, much like the Milne metric of classical GR. With a suitable description of matter properly fitted in, Nash theory can be potentially far more interesting. 

\section{Nash Curvature Correction as a Scalar Field} \label{vacuum_dynamics}
The exact solution portrays a cosmic acceleration and tells us that Nash theory accommodates more solutions compared to GR, even in vacuum. This is due to a geometric source of energy which finds its origin in the higher curvature terms. From what we have seen from the parameter estimation in the last section, this geometric source may not exactly replicate a desired dark energy evolution during the present epoch in Nash theory. However, the fact that it produces an acceleration even without any matter is a notion worthy of further discussions. A few parallels of standard cosmic acceleration can be found in other relevant gravitational systems carrying quadratic curvature terms. For example, it was proved that theories with only a quadratic action can admit \textit{vacuum of arbitrary constant curvature} \cite{deser2}, and a mixed Einstein-quadratic curvature action can generate the effect of a unique cosmological constant term, sometimes termed a \textit{unique vacuum} \cite{deser3}. Given the fact that a Nash action has no Einstein-Hilbert term, we compare the acceleration with a \textit{residual vacuum dynamics} generated from a unique, self-interacting scalar field. We write the Nash Eqs. (\ref{nash1}) and (\ref{nash2}) in the following manner,

\begin{eqnarray}
&& 3H^{2} - 3H^{2} - 3\beta\left(2\ddot{H}H + 6\dot{H}H^{2} - \dot{H}^{2}\right) = 0, \\&&\nonumber
-2\dot{H} - 3H^{2} + (2\dot{H} + 3H^{2}) + \beta\Big(2\overset{\ldots}{H} + 12\ddot{H}H \\&& 
+ 9\dot{H}^{2} + 18\dot{H}H^{2}\Big) = 0.
\end{eqnarray}

Here, $\beta$ is a parameter put in by hand which does not alter the Nash equations in vacuum. The role of this parameter shall be clear in the subsequent discussions. We recall that in standard FRW cosmology the $G^{00}$ equation gives $\rho = 3H^{2}$. Therefore in this equation we can identify $3H^{2}$ as an effective matter density of GR and write it as $\rho_{GR}$ and $-2\dot{H} - 3H^{2}$ as $p_{GR}$. Then the the above set of equations can be re-written as

\begin{eqnarray}\nonumber\label{scalar_intro}
&& \rho_{GR} = 3H^{2} = 3\beta\left(\frac{H^{2}}{\beta} + 2\ddot{H}H + 6\dot{H}H^{2} - \dot{H}^{2}\right), \\&&
p_{GR} = -2\dot{H} - 3H^{2} = -\beta\Big(2\overset{\ldots}{H} + 12\ddot{H}H + 9\dot{H}^{2} \\&&\nonumber
 + 18\dot{H}H^{2} + \frac{2\dot{H} + 3H^{2}}{\beta}\Big) = 0.
\end{eqnarray} 

Let us imagine this reorganization in terms of a self-interacting scalar field $\phi$. The scalar field and its interaction, by virtue of this construction, should obey the dynamical equations
\begin{eqnarray}\label{scalar_int}
&& \frac{\dot{\phi}^{2}}{2} + V(\phi) = 3\beta\left(\frac{H^{2}}{\beta} + 2\ddot{H}H + 6\dot{H}H^{2} - \dot{H}^{2}\right), \\&&\nonumber\label{scalar_int1}
\frac{\dot{\phi}^{2}}{2} - V(\phi) = -\beta\Big(2\overset{\ldots}{H} + 12\ddot{H}H + 9\dot{H}^{2} \\&&\nonumber
 + 18\dot{H}H^{2} + \frac{2\dot{H} + 3H^{2}}{\beta}\Big).
\end{eqnarray}
Let us suppose that the scalar self-interaction is given by the $\phi^4$ potential used in spontaneous symmetry breaking. Then it is easy to see that for mathematical consistency the mass term must be time-dependent. We can then write $V(\phi)$ as
\begin{equation}\label{potential}
V(\phi) = V_{0} - M(t)^{2} \phi^{2} + \frac{\lambda}{4} \phi^{4}.
\end{equation}
The scalar field and $M(t)$ both have mass dimension one while $\lambda$ is dimensionless. 
The vacuum expectation value (\textit{vev}) $\textit{v}$ can be derived from this Higgs-type potential as
\begin{align}
\frac{\partial V}{\partial \phi} \bigg\rvert_{\textit{v}} = 0, \,\,\,\,\,\, \textit{v} = \sqrt{\frac{2M(t)^{2}}{\lambda}}.  \label{328}
\end{align}

If we further carry forward the analogy with the Higgs field, this time variation leads to a variation of standard couplings. The fore-bearer of any such ideas allowing variations of fundamental couplings is the \textit{Large Numbers Hypothesis} of Dirac \cite{dirac}, which received many theoretical treatments over time. Phenomenological analysis of these theories are popular and have produced some consequences of considerable interest~\cite{variation}. For example, variation of a fundamental coupling such as the fine structure constant generates non-trivial variations in the gauge and Yukawa couplings~\cite{campbell}, and in turn, the masses of the strongly interacting quarks and gluons through the QCD scale $\Lambda_{QCD}$~\cite{gasser, bagdo}. We speculate, in the present case, that it is possible for the Higgs \textit{vev} to have variations from its present value, during an epoch dominated by higher curvature Nash corrections, which also coincides with an era of electroweak phase transition, in the cosmological past. This variation can lead to quark mass variations but the proton mass will not vary much since it finds its biggest contribution from the QCD scale alone \cite{calmet, lang}. This leads to mild variations in the proton-to-electron mass ratio $\mu \equiv \frac{m_{p}}{m_{e}}$.  \\

The elementary particle masses are proportional to the Higgs vev. If the Yukawa coupling $\lambda_{e,q}$ is a constant, one can write the mass of a quark as
\begin{equation}
m_{e,q} = \lambda_{e,q} v. \label{2-1}
\end{equation}
The proton mass, however, does not depend very much on the Higgs vev. Using a quark mass expansion and the separation of QCD Hamiltonian in gauge-invariant parts, one can write
\begin{equation}
m_{p} = a(\Lambda_{QCD}) + \sum\limits_{q} b_{q} m_{q}. \label{2-2}
\end{equation}

If $\Lambda_{QCD}$ is also a constant (alongwith the Yukawa coupling), from Eqs. (\ref{2-1}) and (\ref{2-2}) we can write
\begin{align}
\frac{\Delta m_{e}}{m_{e}} &= \frac{\Delta v}{v}, \label{2-8} \\
\frac{\Delta m_{p}}{m_{p}} &= \frac{\sum\limits_{q} b_{q} m_{q}}{a(\Lambda_{QCD}) + \sum\limits_{q} b_{q} m_{q}} \frac{\Delta v}{v} = \frac{9}{100} \frac{\Delta v}{v} . \label{2-9}
\end{align}

According to Lattice QCD simulations \cite{gasser} quark masses $\sum\limits_{q} b_{q} m_{q}$ are accountable for less than $10$ percent of the proton mass. Therefore, a variation of Higgs vev alone results in a negligible change in proton mass $\frac{\Delta m_{p}}{m_{p}}$ compared to $\frac{\Delta m_{e}}{m_{e}}$. From Eqs. (\ref{2-8}) and (\ref{2-9})  
\begin{equation}
\frac{\Delta \mu}{\mu} = \frac{\Delta m_{p}}{m_{p}} - \frac{\Delta m_{e}}{m_{e}} =  - \frac{91}{100} \frac{\Delta v}{v}. \label{2-10}
\end{equation}
If $v_{0}$ and $v_{z}$ are the values of Higgs vev at the present epoch (the present value of Higgs vev is $v_{0} = 246 GeV$) and at some redshift $z$, then $\Delta v / v$ is equivalent to $(v_{z}-v_{0}) / v_{0}$. This variation can be measured from the molecular absorption spectra of a series of quasars. From the observational data-sets one can also infer that the $\mu$ variation is directly proportional with a fine structure constant variation, with the proportionality constant being measured as $R \sim -50$ \cite{avelino}, and related to high-energy scales in theories of unification. The theoretical $\mu$ variation and $\textit{v}$-variation are connected through the equation (for more discussion see \cite{sc}). \\ 

If $\textit{v}(z)$ is derived as a function of redshift using Eq.~(\ref{328}), it is possible to calculate $\textit{v}_{0}$, the value at the present epoch, at $z = 0$. Then,
\begin{equation}
\frac{\Delta \textit{v}}{\textit{v}} = \frac{(\textit{v}_{z} - \textit{v}_{0})}{\textit{v}_{0}}.
\end{equation}
 
We compare the theoretically derived $\frac{\Delta \mu}{\mu}$ with the observational bound found from the data analysis of Cesium Atomic Clock spectroscopy~\cite{hunte} 
\begin{equation}
\frac{\Delta{\mu}}{\mu} = (-0.5 \pm 1.6)\times 10^{-16} \, \, year^{-1} .
 \label{017}
\end{equation}
It is more practical to write the variation in comparison with the Hubble parameter whose present value is $H_{0} \simeq 7 \times 10^{-11}  \, \, year^{-1}$. Thus, the variation is in a scale of $\frac{\Delta{\mu}}{\mu} \simeq 10^{-6} H_{0}$. The observed variations in different Redshift are given in Table. \ref{table1}, as weighted average analyses of Hydrogen molecular spectra from different quasars \cite{king, malec, kanekar, weerd, wendt, ubachs, dapra}.

\begin{table}
\caption{{\small $\Delta \mu / \mu$ vs redshift $z$. Weighted average analyses of Hydrogen molecular spectra from different quasars for $z > 2$ alongwith a few other molecular spectra for $z < 1$.}}\label{table1}
\begin{tabular*}{\columnwidth}{@{\extracolsep{\fill}}lrrrrl@{}}
\hline
\multicolumn{1}{c}{Quasar} & \multicolumn{1}{c}{Redshift} & \multicolumn{1}{c}{$\Delta \mu / \mu \, \, [10^{-6}H_0]$} \\
\hline
B0218+357 & 0.685 & $ -0.35  \pm 0.12 $ \\
PKS1830-211 & 0.89 & $ 0.08  \pm 0.47 $ \\
HE0027-1836 & 2.40 & $ -7.6  \pm 10.2 $ \\
Q0347-383 & 3.02 & $ 5.1  \pm 4.5 $ \\
Q0405-443 & 2.59 & $ 7.5  \pm 5.3 $ \\
Q0528-250 & 2.81 & $ -0.5  \pm 2.7 $ \\
B0642-5038 & 2.66 & $ 10.3  \pm 4.6 $ \\
J1237+064 & 2.69 & $ -5.4  \pm 7.2 $ \\
J1443+2724 & 4.22 & $ -9.5  \pm 7.5 $ \\
J2123-005 & 2.05 & $ 7.6  \pm 3.5 $ \\
\hline
\end{tabular*}
\end{table}

\begin{figure}
\begin{center}
\includegraphics[width=0.40\textwidth]{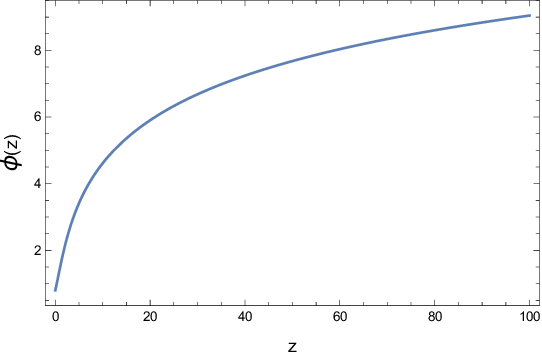}
\includegraphics[width=0.40\textwidth]{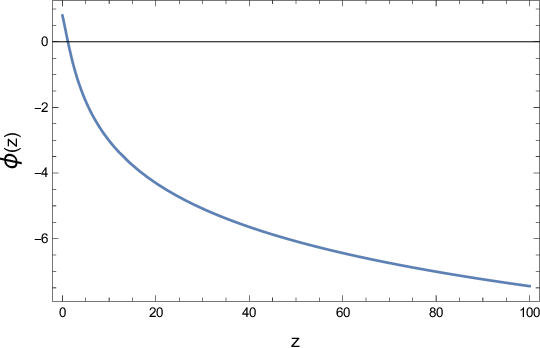}
\caption{Evolution of the conjectured vacuum scalar field $\phi$ as a function of redshift.}
\label{plot_0}
\end{center}
\end{figure}

\begin{table*}[]
\begin{center}
\begin{tabular}{|c|c|c|c|}
\hline
 Source & Redshift & ${\Delta\alpha}/{\alpha}$ (ppm) & Spectrograph. \\
\hline\hline
J0026$-$2857 & 1.02 & $3.5\pm8.9$ & UVES \\
\hline
J0058$+$0041 & 1.07 & $-1.4\pm7.2$ & HIRES \\
\hline
3 sources & 1.08 & $4.3\pm3.4$ & HIRES \\
\hline
HS1549$+$1919 & 1.14 & $-7.5\pm5.5$ & UVES/HIRES/HDS \\
\hline
HE0515$-$4414 & 1.15 & $-1.4\pm0.9$ & UVES \\
\hline
J1237$+$0106 & 1.31 & $-4.5\pm8.7$ & HIRES \\
\hline
HS1549$+$1919 & 1.34 & $-0.7\pm6.6$ & UVES/HIRES/HDS \\
\hline
J0841$+$0312 & 1.34 & $3.0\pm4.0$ & HIRES \\
J0841$+$0312 & 1.34 & $5.7\pm4.7$ & UVES \\
\hline
J0108$-$0037 & 1.37 & $-8.4\pm7.3$ & UVES \\
\hline
HE0001$-$2340 & 1.58 & $-1.5\pm2.6$ &  UVES \\
\hline
J1029$+$1039 & 1.62 & $-1.7\pm10.1$ & HIRES  \\
\hline
HE1104$-$1805 & 1.66 & $-4.7\pm5.3$ & HIRES \\
\hline
HE2217$-$2818 & 1.69 & $1.3\pm2.6$ &  UVES \\
\hline
HS1946$+$7658 & 1.74 & $-7.9\pm6.2$ & HIRES \\
\hline
HS1549$+$1919 & 1.80 & $-6.4\pm7.2$ & UVES/HIRES/HDS \\
\hline
Q1103$-$2645 & 1.84 & $3.5\pm2.5$ &  UVES \\
\hline
Q2206$-$1958 & 1.92 & $-4.6\pm6.4$ &  UVES \\
\hline
Q1755$+$57 & 1.97 & $4.7\pm4.7$ & HIRES  \\
\hline
PHL957 & 2.31 & $-0.7\pm6.8$ & HIRES  \\
PHL957 & 2.31 & $-0.2\pm12.9$ & UVES  \\
\hline
\end{tabular}
\caption{\label{table2} The data-table for the variation of ${\Delta\alpha}/{\alpha}$. Unit of variation is parts per million (ppm).}
\end{center}
\end{table*}

\begin{figure}
\begin{center}
\includegraphics[width=0.40\textwidth]{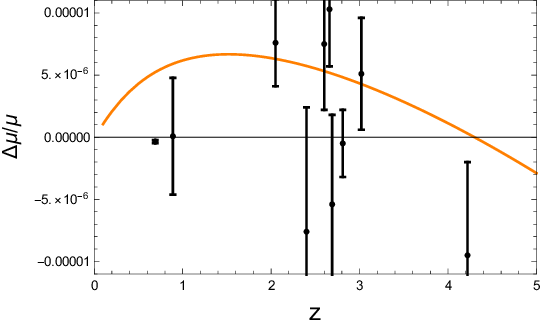}
\includegraphics[width=0.40\textwidth]{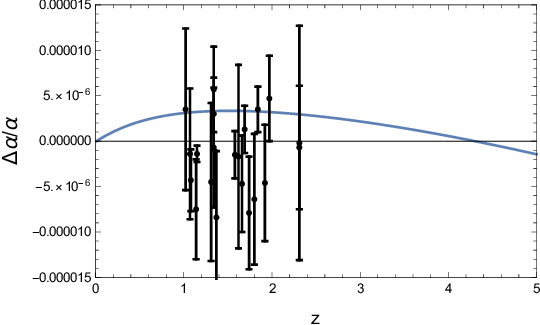}
\caption{Top Panel : Evolution of the conjectured $\Delta \mu / \mu $ as a function of redshift alongwith the fitted observational points from quasar absorption spectra, following Table \ref{table1}. The associated error bars are also shown in the graph. Bottom Panel : Evolution of the conjectured $\Delta \alpha / \alpha $ as a function of redshift alongwith the fitted observational points from quasar absorption spectra, following Table \ref{table2}. The associated error bars are also shown in the graph.}
\label{plot_1}
\end{center}
\end{figure}

While the data in Table \ref{table1} clearly shows a variation of the coupling during cosmic expansion, one canot claim anything about its mathematical form. In the case of Nash vacuum, we have actually solved the scalar field equation Eq.~(\ref{scalar_int}) to derive a desired form. For convenience, we express $M(t)$ as a function of redshift and write
\begin{equation}
M(t)^{2} = M_{0}^{2} u(z).
\end{equation}  
The parameter $M_{0}$ is of mass dimension one in natural units. The dimensionless function of redshift $u(z)$ holds the key for any variation of $\mu$. Next, using Eqs. (\ref{scalar_int}) and (\ref{scalar_int1}) we write
\begin{equation}\label{tosolve}
\dot{\phi} = \pm \left\lbrace-2\dot{H} - \beta\left(12\dot{H}^{2} + 6\ddot{H}H + 2\overset{\ldots}{H}\right)\right\rbrace^{\frac{1}{2}},
\end{equation}
and transform the equation into arguments of redshift, using Eq. (\ref{hubble}). The numerical solutions of the transformed equation are plotted in Fig. \ref{plot_0}. The two curves in Fig. \ref{plot_0} are for the two different signs, plus and minus on the RHS of Eq. (\ref{tosolve}). Using the solution for the scalar field we solve for $u(z)$, or $M(t)^{2}$, and estimate the Higgs \textit{vev} $\textit{v}(z)$. Then using $\textit{v}(z)$, we find $\frac{\Delta \mu}{\mu}(z)$ and plot it in the top panel of Fig. \ref{plot_1}. The data points from Table. \ref{table1} are also fitted with the curve. Although the curve does not exactly move through all the data points, at a scale of variation given by $\simeq 10^{-6} H_{0}$ it is not expected to do so either. The curve falls within the overall range of variation taken from observations of Molecular absorption spectra, only if $0.001 < \beta < 0.01$. The plot in Fig. \ref{plot_1} is for $\beta = 0.005$. In our interpretation it is the best fit parameter value of $\beta$, although no proper estimation technique is involved here apart from a classic trial and error method. This mild cosmic variation and its comparison with molecular absorption lines of Quasar spectra also provides us motivation to look for variations in fine structure constant $\alpha$. Once again, we refer to the data set from a combined analysis of constraints that uses molecular absorption spectroscopy of different Quasars \cite{webb, ferreira, martins, whitmore, pinho}. The variations are quite mild in the scale of $\simeq 10^{-6} H_{0}$ or parts per million (ppm) and as expected, more mild compared to the $\mu$-variation. The specific set of measurements of $(\alpha_{z}-\alpha_{0}) / \alpha_{0} = \Delta \alpha / \alpha$ taken here are from the HIRES and UVES spectrographs \cite{aga, reimers, molaro, evans}, operated at the Keck and VLT telescopes. The measurements from a variety of source Quasars are written in the form of Table. \ref{table2}. We fit these measurements with the derived evolution of $\Delta\alpha/\alpha$ and give the combined curve in the bottom panel of Fig. \ref{plot_1}. The scale of variation is shown for $\beta = 0.005$. The parameter $\beta$ therefore seems to be a necessary coupling of the theory. It can be called a weight factor, that serves our requirement of observational validity. Looking into Eq. (\ref{scalar_intro}), it is evident that $\beta$ should have a dimension $-2$ in natural units ($M_{P}^{-2}$). At this point, we make a speculation that $\beta \sim H_{0}^{-2}$ and leave this particular issue to future discussions. We also note that the derived variation of $\Delta \mu / \mu$ and $\Delta \alpha / \alpha$ seems like one half of a sinusoid and inspires speculations regarding a probable oscillatory behavior of the variation, within proper scale. Similar variations have recently been reported in the context of generalized Brans-Dicke type theories \cite{sc} and models of running vacuum \cite{sola}.  

\section{Conclusions} \label{conclusions}
Quite a few quadratic actions of gravity have been considered in gravitational physics and their implications in different issues, related to renormalizability or cosmic expansion history are more or less well documented. A Nash theory appears to be a specific choice within a larger class of theories \cite{Stelle}, that promote field equations of higher order and remains relatively unexplored. Due to the specific comparative weighting of the term quadratic in the scalar curvature and that quadratic in the Ricci tensor in the action, the field equations of Nash theory provides a few unique properties. The primary motivation is to look into these properties from a cosmological perspective, even though the original theory was just conceived out of a mathematical curiosity : to write \textit{`an interesting equation'} for a supposedly Yukawa-like gravitational field that can relate to Klein-Gordon equations in a non-relativistic context. The additional purpose of this manuscript is to reaffirm that a Nash equation, the equation with a Yukawa-like aspect can indeed be portrayed as an \textit{alternative vacuum equation} \cite{Nash_lec}. While any solution of GR is automatically a solution of a Nash theory, the reverse is not always true. One might find one or two solutions or properties which are unique to the Nash dynamics but not found in GR. For instance, exact solution of static Nash field equations for empty space produces a Schwarzschild-de-Sitter spacetime. Moreover, the simplest non-trivial solution of the field equations in the cosmological case is simply, the de-Sitter spacetime [ref]. Hence empty space in the Nash theory naturally corresponds to a space with Lorentz Invariant Vacuum Energy and perhaps makes dark energy superfluous. Moreover, as pointed out by Nash himself, the scalar equation of the theory can easily be derived from the vacuum tensor equation and for $4D$ it leads simply to $\square R = 0$, a form extremely suggestive of waves. It asserts simply that the scalar curvature satisfies the wave operator in $3 + 1$ dimensions. These remain as the curiously distinct properties of a Nash theory. \\

We work with the spatially flat cosmological equations in the Nash theory. No source of matter, fluid, exotic fields or a cosmological constant is taken. The field equations, although carrying fourth order derivatives of the scale factor, due their characteristic symmetry, is governed by a single independent component in the form of Eq. (\ref{vaccnash1}). The component leads to a simple yet non-trivial relation between the kinematic parameters jerk and deceleration in the form of Eq. (\ref{jqrelation}) which is a distinct property of Nash cosmology. It ensures that the cosmological evolution is constrained at the outset such that one may not choose these parameters at will. We solve the field equations directly to write an exact solution for the Hubble parameter. The viability of this solution is discussed in comparison with a wide array of observational data from the present cosmological epoch. The Hubble evolution is satisfactory compared to the luminosity distance modulus measurements and the evolution of deceleration parameter suggests an epoch of deceleration prior to the present acceleration. The redshift of transition is found to be quite satisfactory. There are, however, a few departures. The present value of deceleration does not match exactly with standard $\Lambda$CDM cosmology; the effective EOS of the system suggests a radiation dominated era of deceleration rather than a matter-dominated one. These issues can indeed be addressed, gradually, once we figure out the best possible way to include a energy momentum tensor within the theory to engineer some fine-tunings in this otherwise crude yet promising cosmological behavior.  \\

To portray the Nash equations as a modified vacuum equation, we incorporate a Higgs-like scalar field. We manipulate the vacuum equations such that they seem equivalent to a standard GR vacuum plus the Higgs scalar and its self-interaction. We find that the vacuum expectation value of this so-called Higgs scalar field should be a function of coordinates (time alone in this case) in order for this construction to be consistent. This opens up an interesting discussion if we identify this scalar as the Standard Model Higgs field. In such a scenario, this variation leads to non-trivial variations in proton-electron mass ratio and the fine structure constant, fundamental couplings which are otherwise deemed to be constants in the standard model of particle physics. Moreover, the scale of this variation matches quite well with the analyzed data of molecular absorption spectra from a number of Quasars. If we expect the vacuum energy density to be a dynamical quantity in an expanding universe, according to quantum field theory of curved space-time it should exhibit a slow evolution, determined by the expansion rate of the universe. Recent measurements on the time variation of the fine structure constant and of the proton-to-electron mass ratio suggest that basic quantities of the Standard Model, such as the either or both of the QCD scale parameter and Higgs Vacuum Expectation Value may not be conserved in the course of the cosmological evolution \cite{sola111}. The masses of the nucleons and of the atomic nuclei would also be affected and Matter is not conserved in such a universe \cite{sola112}. These measurements can be interpreted as a leakage of matter into vacuum or vice versa \cite{sola113, sola114}. We want to point out that this variation can be a signature of how gravity behaves in a common platform with the standard model of particle physics. While the variation may have a possible contribution from the dark matter particles, interaction of the so-called dark energy with ordinary baryonic matter as well as with dark matter, our proposal is that it also depends on the higher curvature corrections, such as the Nash action. One might also derive the scale of variation for a standard $R^2$ theory, however, the pattern of variation found their should be different. At this moment, the theory of Nash is treated simply as a theory of vacuum alone and can emphasize the recent interests in dealing with vacuum dynamical solutions (see for instance, the work of Moreno-Pulido and Sola~\cite{sola01, sola02} and the monograph by Sola \cite{sola03} for a review on recent developments on running vacuum). In all of these cases, the vacuum is depicted as a solution embedded in GR, generating dynamical evolution of the vacuum energy. As the present manuscript suggests, a similar phenomenology can also be generated by adding higher curvature terms (non-lovelock) into the Einstein-Hilbert action. \\

We conclude with a positive intent and curiosity, by commenting that it is possible for the Nash theory to emerge as a good theory and provide more of such interesting phenomenology. A simple vacuum solution of the theory can challenge our usual understandings of the nature, for instance, the endurance of a \textit{fundamental constant} approach or the viability of the Equivalence principles. We need to formulate a stronger version of the theory that can fill in different gaps. One possible way of doing that is perhaps already hinted in this manuscript, through the weight parameter $\beta$ which needs to be within a certain range to minimize the variation of fundamental couplings. One may imagine this parameter to be a different scalar field of the dimension $H_{0}^{-2}$ and formulate a Scalar-Nash theory. We keep in mind that any solution of an Einstein vacuum is already a solution of Nash vacuum, however, this never excludes the possibility of finding a larger class of new solutions. Therefore, another way to proceed might be to look for new static solutions in the theory and check for viability of standard principles that are viable in GR, such as the Birkhoff Theorem. These will be addressed in near future by the authors. 

\section{Acknowledgements}

Two of the authors (RB and SB) acknowledge the support from a DAE Raja Ramanna Fellowship (grant no: $1003/(6)/2021/RRF/R\&D-II/4031$, dated: $20/03/2021$). SC acknowledges Vellore Institute of Technology for the financial support through its Seed Grant (No. SG20230027), year 2023.

\end{document}